# Detectability of biosignatures in anoxic atmospheres with the James Webb Space Telescope: A TRAPPIST-1e case study

Joshua Krissansen-Totton[1]*, Ryan Garland[2], Patrick Irwin[2], David C. Catling[1]
[1]Department of Earth and Space Sciences/Astrobiology Program, Johnson Hall, University of Washington, Seattle, WA, 98195, USA.
[2]Atmospheric, Ocean, and Planetary Physics, Clarendon Laboratory, Department of Physics, University of Oxford, Oxford OX1 3PU, UK.
*Corresponding author: joshkt@uw.edu

Short Title: Detectability of biosignatures with JWST

## Abstract

The James Webb Space Telescope (JWST) may be capable of finding biogenic gases in the atmospheres of habitable exoplanets around low mass stars. Considerable attention has been given to the detectability of biogenic oxygen, which could be found using an ozone proxy, but ozone detection with JWST will be extremely challenging, even for the most favorable targets. Here, we investigate the detectability of biosignatures in anoxic atmospheres analogous to those that likely existed on the early Earth. Arguably, such anoxic biosignatures could be more prevalent than oxygen biosignatures if life exists elsewhere. Specifically, we simulate JWST retrievals of TRAPPIST-1e to determine whether the methane plus carbon dioxide disequilibrium biosignature pair is detectable in transit transmission. We find that ~10 transits using the Near InfraRed Spectrograph (NIRSpec) prism instrument may be sufficient to detect carbon dioxide and constrain methane abundances sufficiently well to rule out known, non-biological $CH_4$ production scenarios to ~90% confidence. Furthermore, it might be possible to put an upper limit on carbon monoxide abundances that would help rule out non-biological methane-production scenarios, assuming the surface biosphere would efficiently drawdown atmospheric CO. Our results are relatively insensitive to high altitude clouds and instrument noise floor assumptions, although stellar heterogeneity and variability may present challenges.

## 1) Introduction

The James Webb Space Telescope (JWST) will provide the first opportunity to look for non-intelligent life beyond the solar system. To date, much of the focus on exoplanet biosignatures has been on molecular oxygen (Brandt & Spiegel 2014; Owen 1980) and its photochemical product ozone (e.g. Domagal-Goldman et al. 2014; Harman et al. 2015; Leger et al. 1993; Meadows et al. 2018b; Segura et al. 2003; Tian et al. 2014). Oxygen is an excellent biosignature gas because it is challenging to produce in large quantities without life, and although hypothetical false-positives scenarios have been proposed, a growing understanding exists of how they might be distinguished using contextual clues (Harman & Domagal-Goldman 2018; Meadows 2017; Meadows et al. 2018b; Schwieterman et al. 2016).

However, even if life is widespread in the cosmos there is no guarantee that oxygen-rich biosignatures are common. Oxygenic photosynthesis is an extremely complex metabolism that only evolved once in Earth history (Knoll 2008; Mulkidjanian et al. 2006), and the emergence of oxygenic photosynthesis does not necessarily guarantee an oxygen-rich atmosphere because biogenic oxygen sources may be overwhelmed by non-biological sinks (Catling & Claire 2005). Oxygen concentrations may stabilize at low, but difficult to detect levels (Reinhard et al. 2017), and the accumulation of oxygen on planets around M-dwarfs may be especially difficult due to the low flux of visible photons (Lehmer et al. 2018).

These limitations led Krissansen-Totton et al. (2018) to propose the combination of carbon dioxide ($CO_2$) and abundant biogenic methane ($CH_4$) as an alternative biosignature for anoxic atmospheres. Carbon dioxide and methane would have been out of chemical equilibrium on the early Earth during the Archean eon (4.0-2.5 Ga), and their coexistence cannot be explained without a continuous replenishing flux of $CH_4$, which demands a biological source. Specifically, for terrestrial planets, > 0.1% methane abundance is potentially biogenic and > 1% is probably biogenic due to the low likelihood of producing such large quantities of methane through relatively inefficient non-biological processes such as serpentinization followed by Fischer-Tropsch-type reactions (Krissansen-Totton et al. 2018). Furthermore, the inference of biology could be strengthened by the non-detection of carbon monoxide (CO) because several non-biological scenarios that produce $CH_4$ would also be expected to produce CO (Krissansen-Totton et al. 2018).

The ability of JWST to characterize terrestrial planets and detect oxygen-ozone biosignatures has been studied extensively (Barstow & Irwin 2016; Deming et al. 2009; Greene et al. 2016; Kaltenegger & Traub 2009; Morley et al. 2017). Barstow et al. (2015) modeled the modern Earth transiting an M6V star and found that a year of primary transits might be sufficient for a tentative detection of $CO_2$ and $O_3$. Irwin et al. (2014) also demonstrated that atmospheric characterization of Earth-like planets around M-dwarfs is possible. Barstow and Irwin (2016) applied these calculations specifically to the TRAPPIST-1 system (Gillon et al. 2017) and found that for modern Earth-like atmospheres 30-60 transits would be necessary to reliably detect Earth-like $O_3$ levels on 1b, 1c, and 1d. The outer planets (1e, 1f, 1g) are more likely to be habitable (Turbet et al. 2018; Wolf 2017), but $O_3$ detection would require even more transits in these cases (Barstow & Irwin 2016). These detections are barely feasible given the expected lifetime of the JWST mission and the position of TRAPPIST-1 in the sky close to the ecliptic, which limits the star's viewing by JWST.

The retrieval studies cited above used optimal estimation techniques and assumed photon limited noise with some allowance for instrument throughput, but others have performed Markov Chain Monte Carlo (MCMC) retrievals with more realistic instrument simulators. Greene et al. (2016) simulated a MCMC retrieval on a single transit of a cool (500 K) super Earth and found that $H_2O$ and $CH_4$ detections might be possible in cloud and haze free atmospheres. Morley et al. (2017) calculated the number of primary transits required to rule out a flat spectrum to 5σ for TRAPPIST-1 planets and found relatively few (10 or less) would be needed for TRAPPIST-1e. Batalha et al. (2018) performed an information analysis to show that a partial saturation strategy using the NIRSpec instrument could plausibly constrain the atmospheric composition of terrestrial planets. However, the detectability of the $CH_4$+$CO_2$ biosignature combination has not been investigated.

Here, we simulated retrievals to investigate the detectability of $CO_2$+$CH_4$ biosignatures with JWST. The relatively primitive nature of methanogenesis (Weiss et al. 2016) and its antiquity on Earth (Wolfe & Fournier 2018) suggests that this anoxic biosignature is probably more common than oxygen-ozone biosignatures (Krissansen-Totton et al. 2018). Here, we demonstrate that it is also more easily detectable than ozone in transit transmission observations with JWST. We focus on TRAPPIST-1e because climate models suggest its surface conditions are potentially habitable for a wide range of atmospheric compositions (Turbet et al. 2018; Wolf 2017) and because it is the only TRAPPIST-1 planet with a bulk composition consistent with an Earth-like iron core (Suissa & Kipping 2018). However, our results apply similarly to TRAPPIST-1f and 1g, and other nearby, habitable planets that will be discovered by the Transiting Exoplanet Survey Satellite (TESS) mission (Kempton et al. 2018). Finally, we present calculations showing how JWST detections of $CH_4$ and $CO_2$ might be used to quantitatively evaluate the likelihood of a surface biosphere.

## 2) Methods

Transmission spectra for TRAPPIST-1 planets were calculated using the NEMESIS radiative transfer code (Irwin et al. 2008). NEMESIS uses a correlated-k radiative transfer scheme (Goody & Yung 1995; Lacis & Oinas 1991), and in this study k-tables were calculated from the HITRAN 2008 line database (Rothman et al. 2009), except for methane Near Infrared (NIR) band data, which was taken from Karkoschka and Tomasko (2010). The k-tables were calculated at a spectral resolution of 0.025 μm at 20 temperatures in the range 70-400 K, and 20 pressures equally spaced in log space from $3.1 \times 10^{-7}$ to 20.3 bar (see Irwin et al. (2014) for further details).

We generated synthetic spectra for TRAPPIST-1e by adopting recent mass (Grimm et al. 2018) and radius (Delrez et al. 2018) estimates and assuming an atmospheric composition. We initially assume a 1-bar Archean-like atmospheric composition dominated by $N_2$, with 5% $CO_2$, 0.5% $CH_4$, and 10 ppb CO, where all gas mixing ratios are constant with altitude. Our chosen methane abundance is representative of plausible biogenic methane fluxes on the early Earth (Kharecha et al. 2005), whereas the high $CO_2$ abundance was chosen to ensure a habitable surface climate (Turbet et al. 2018; Wolf 2017) whilst avoiding significant haze-formation (Arney et al. 2017), which we do not include in our nominal retrieval. We also considered a modern Earth analog of TRAPPIST-1e with 20% $O_2$, 290 ppm $CO_2$, 1.7 ppm $CH_4$, and 0.1 ppm CO. To be generous for ozone retrieval (since we will argue that ozone retrieval is difficult) we assumed a very high estimate of 0.01% $O_3$. This concentration is about ~10 times higher than the peak ozone abundance of 10 ppmv in the modern Earth's stratosphere at ~25 km altitude. Our assumed $O_3$ concentration is also ~10x higher than the predicted peak ozone abundance for an Earth-like Proxima Centauri b calculated self-consistently using a photochemical model and assuming Earth-like biological fluxes (Meadows et al. 2018a).

For synthetic spectra, assumptions were also made about atmospheric structure and water vapor. Both atmospheres were assumed to be isothermal above 0.1 bar (Robinson & Catling 2014), and follow a moist adiabatic lapse rate below 0.1 bar with a smoothed transition between the two regions. The stratospheric temperature was assumed to be 214 K, which is the skin temperature of the modern Earth. The atmospheric temperature profile was not calculated using a climate model, but this is unlikely to affect our results significantly because transmission spectra are relatively insensitive to atmospheric temperature structure. For both the Archean Earth and modern Earth cases, atmospheric water vapor was assumed to be 1%, or water vapor saturation (whichever is smaller). This vastly overestimates stratospheric water vapor content because the cold trap limits stratospheric water vapor mixing ratios to a few ppmv on the modern Earth (Oman et al. 2008) and $<10^{-6}$ for early Earth-like atmospheres (Kasting & Ackerman 1986). However, 1% water vapor represents a pessimistic end-member assumption about the extent to which water vapor absorption features could obscure those of $CH_4$ and $CO_2$. Sensitivity tests with ppmv stratospheric water vapor show that posterior uncertainties in $CH_4$ and $CO_2$ are decreased by lowering stratospheric H$_2$O (not shown). Our stellar spectrum for TRAPPIST-1 was identical to that adopted in Barstow and Irwin (2016).

Initially, we assumed all atmospheres were cloud free, but later tested a grey, single layer cloud model. In this model, clouds are described by three parameters: nadir optical depth, cloud base pressure, and fractional scale height. If parameters are chosen to represent Earth-like water clouds (e.g. Irwin et al. 2014) the transit spectrum is truncated at around ~20 km (0.05 bar) above the surface, which we found has a minimal impact on the retrieval. Instead, we chose cloud parameters to truncate the transmission spectrum at ~30 km (0.01 bar) to approximate opacity due to very high altitude clouds, an organic haze (e.g. Arney et al. 2017), or sulfate aerosols (Misra et al. 2015).

To add realistic observational noise to our synthetic spectra, we used the JWST instrument simulator PandExo (Batalha et al. 2017). NIRSpec prism was used to simulate Archean Earth-like spectra because its 0.6-5.3 μm range allows simultaneous coverage of CH$_4$, CO$_2$, and CO absorption features, whereas both NIRSpec prism and the Mid InfraRed Instrument (MIRI) Low Resolution Spectrometer (LRS) were used to simulate modern Earth-like spectra because MIRI's 5-12 μm range includes the 9.6 um ozone band. Additionally, we adopted the partial saturation strategy described in Batalha et al. (2018) for NIRSpec prism to increase the observing efficiency from 33% to 72%. Unless stated otherwise, we assume zero noise floor, equal time in and out of transit, and 80% saturation level (for MIRI LRS). For convenience, we also binned NIRSpec and MIRI spectra to constant-width bins equal to the size of the largest resolution element in each instrument. This results in some information loss but is unlikely to significantly impact our retrievals. Typically, we will add random noise instances to the true spectrum to generate synthetic spectra for retrieval. However, we sometimes place the midpoint of all data points on the true spectrum to ensure posteriors are centered on true values and not biased by a handful of data points. Feng et al. (2018) demonstrated that these centered posteriors are essentially identical to the summation of posteriors from many individual noise realizations.

To solve the inverse problem and retrieve planet parameters we used the Nested Sampling algorithm (Feroz & Hobson 2008; Feroz et al. 2009) implemented using PyMultiNest (Buchner et al. 2014). Nested Sampling is a Bayesian retrieval algorithm that samples equal-likelihood regions of prior-space to explicitly calculate the Bayesian evidence, the denominator in Bayes' theorem. Posterior probability distributions for unknown parameters can then be calculated from the Bayesian evidence (Feroz et al. 2009). We compared posteriors to those from emcee (Foreman-Mackey et al. 2013) and they were virtually identical to Nested Sampling posteriors for retrievals with the same priors and likelihood function. Table 1 shows the input parameter values and their uniform priors for the Archean Earth-like and modern Earth-like spectra.

**Table 1**: Assumed parameter values used to create synthetic spectra, and uniform prior ranges adopted for simulated retrieval.

|  | Archean Earth-like TRAPPIST-1e | | Modern Earth-like TRAPPIST-1e | | Archean Earth-like TRAPPIST-1e with clouds | |
|---|---|---|---|---|---|---|
|  | Assumed value | Uniform prior | Assumed value | Uniform prior | Assumed value | Uniform prior |
| Log(CH$_4$) | -2.3010 (0.5%) | [-8.0,0.0] | -5.77 (1.7 ppm) | [-8.0,0.0] | -2.3010 (0.5%) | [-8.0,0.0] |
| Log($CO_2$) | -1.3010 (5%) | [-8.0,0.0] | -3.5367 (290 ppm) | [-8.0,0.0] | -1.3010 (5%) | [-8.0,0.0] |
| Log(CO) | -8.0 (10 ppb) | [-8.0,0.0] | -6.91 (0.1 ppm) | [-8.0,0.0] | -8.0 (10 ppb) | [-8.0,0.0] |
| Log ($H_2O$) | -2.0 (1%) | [-8.0,0.0] | -2.0 (1%) | [-8.0,0.0] | -2.0 (1%) | [-8.0,0.0] |
| Log($O_2$) | N/A | N/A | -0.69897 (20%) | [-8.0,0.0] | N/A | N/A |
| Log($O_3$) | N/A | N/A | -4.0 (0.01%) | [-8.0,0.0] | N/A | N/A |
| Radius [$R_{Earth}$] | 0.91 | [0.8,1.1] | 0.91 | [0.8,1.1] | 0.91 | [0.8,1.1] |
| Mass* [$M_{Earth}$] | 0.772 | σ=0.077* | 0.772 | σ=0.077* | 0.772 | σ=0.077* |
| $P_{surf}$ (log (bar)) | 0.0 | [-3,2] | 0.0 | [-3,2] | 0.0 | [-3,2] |
| $T_{strat}$ (K) | 214.4 | [100,400] | 214.4 | [100,400] | 214.4 | [100,400] |
| $P_{cloud-base}$ (log(bar)) | N/A | N/A | N/A | N/A | -0.48148 (0.3 bar) | [-3,2]** |
| Optical depth (log(τ)) | N/A | N/A | N/A | N/A | 1 (10) | [-10,5] |
| Fractional scale height, log(FSH) | N/A | N/A | N/A | N/A | -0.5 (0.32) | [-2,1] |

*Rather than use a uniform prior for planet mass, we adopt the mass distribution obtained from transit timing variations in Grimm et al. (2018). This distribution is accurately approximated by a Gaussian with a mean of 0.772 $M_{Earth}$ and σ=0.077 $M_{Earth}$. Note that the posterior distribution for planet mass is nearly identical to this prior because mass is not constrained by transit observations.

**Cloud base pressure is constrained to always be smaller than the surface pressure.

**3) Results**

Fig. 1a and 1b show the Archean Earth-like NIRSpec prism transmission spectrum for TRAPPIST-1e generated using PandExo. The median fitted spectrum from the Nested Sampling retrieval with 95% credible intervals are also plotted. Fig. 2 shows the corresponding posterior probability distributions for the 8 model parameters defined in Table 1. The assumed input ("true") parameter values are over-plotted as vertical and horizontal blue lines on these posteriors. For this 10-transit case, both $CO_2$ and $CH_4$ are detectable, and it is possible to constrain the $CH_4$ abundance (log($CH_4$) = $-2.23^{+0.78}_{-0.96}$). Additionally, a tentative upper limit can be placed on CO abundance (CO < 652 ppm with 90% credibility). Surface pressure cannot be well-constrained, which contributes to the uncertainty in mixing ratio abundances because absorption features can be explained by high abundances and low total pressure, or low abundances and high total pressure (Benneke & Seager 2012). This degeneracy can be seen in the joint distributions in Fig. 2 where there is a negative correlation between gas mixing ratios (particularly $CH_4$ and $CO_2$) and surface pressure. There is a related degeneracy between planet radius and surface pressure because smaller planet radii must be offset by a large surface pressure to fit the same transit depth. Note that our radius parameter is defined as the solid body radius. In Appendix B we repeat the retrieval defining radius as the 1 mbar planet radius. This alternative formulation produces joint distributions for gas abundances and radius that are tightly anticorrelated, but the marginal distributions for gas abundances are unchanged.

Transit spectroscopy does not provide strong constraints on atmospheric temperature structure, but the stratospheric temperature posterior in Fig. 2 is constrained by our tight prior on planet mass from Grimm et al. (2018). This prior for planet mass breaks the degeneracy between mass and stratospheric temperature. Without it, these two parameters would be strongly positively correlated because atmospheric scale height is proportional to temperature/gravity.

Fig. 1c and 1d show the modern Earth-like transmission spectrum for TRAPPIST-1e from NIRSpec prism, whereas Fig. 1e and 1f show the modern Earth-like transmission spectrum for TRAPPIST-1e from MIRI LRS. Fig. 3 shows a comparison between the posterior distributions for methane abundance from the Archean Earth-like case (Fig. 3a), the ozone abundances the modern Earth-like case using NIRSpec prism (Fig. 3b), and the modern Earth-like case using MIRI LRS (Fig. 3c). Since constraining $CH_4$ abundance is crucial for determining the biogenicity of $CH_4$-$CO_2$ disequilibria, and ozone is the most easily observable biosignature gas for modern Earth-like atmospheres, this figure directly contrasts the detectability of Archean-Earth and modern-Earth biosignatures. Whereas it is possible to constrain $CH_4$ abundances to within 1-2 orders of magnitude with 10 transits with NIRSpec prism, even high assumed $O_3$ typically cannot be detected with 10 transits with either NIRSpec prism or MIRI LRS. This is because the uncertainties in transit depth are much larger around prominent ozone features than they are around $CH_4$ absorption features (Fig. 1).

Fig. 4 shows how the uncertainty in key model parameters changes as the number of co-added transits is increased (Archean Earth-like case). Increasing the number of transits reduces the uncertainty in almost all model parameters, but there are diminishing returns beyond 10 transits, consistent with Batalha et al. (2018). However, 30-50 transits would reduce the 66% credible interval in $CH_4$ abundance by ~0.5 log unit (and similarly for the 95% credible interval). Tighter constraints on methane abundances would enable stronger inferences to life (see below).

All the results described above are for cloud-free atmospheres. Fig. 5 shows selected posterior probability distributions for our cloudy Archean case. Here, the "true" cloud parameters were chosen to truncate the spectrum at ~30 km (0.01 bar), which is significantly cloudier than the modern Earth where high clouds truncate the transmission spectrum at around 20 km (Irwin et al. 2014). Nonetheless, both $CO_2$ and $CH_4$ are detectable, and $CH_4$ abundances are still constrained ($\log(CH_4) = -2.57^{+0.90}_{-1.09}$), albeit less tightly than the no cloud case.

In summary, these simulated retrievals suggest that the $CH_4$-$CO_2$ disequilibrium biosignature is detectable for Archean Earth-like planets with JWST in ~10 transits. Additionally, it may be possible to place an upper bound on CO to help rule out non-biological scenarios (see discussion for further consideration of non-biological CO production). This biosignature combination should be easier to detect than oxygen or ozone biosignatures with JWST, and the presence of Earth-like clouds should not impede the retrieval.

**4) Discussion**

The results reported here are broadly consistent with those of Greene et al. (2016), who performed MCMC retrievals for simulated JWST transit transmission observations of a cloud-free, 500 K super-Earth with a $CH_4$ mixing ratio of $4.3\times10^{-4}$ and negligible $CO_2$ and CO. Their posterior distribution for $CH_4$ extends across 2-3 log units, in agreement with our Fig. 2. Additionally, their retrieved upper limit for CO is consistent with our Fig. 2, although they were unable to constrain the $CO_2$ mixing ratio. Note however, that Greene et al. (2016) only considered a single transit, their assumed planet-to-star radius was less favorable than for the TRAPPIST-1 system, and they combined a different suite of instruments, and so some differences are expected. Our results are also broadly consistent with those of Morley et al. (2017) who calculate that <10 transits are required to rule out a flat spectrum at 5σ confidence for TRAPPIST-1e with a $CO_2$-rich atmosphere.

Ultimately, we would like to use gas abundance constraints from JWST observations to evaluate the probability of a planet hosting life. One possible approach is to convert the methane abundance posterior to a probability distribution for the required surface $CH_4$ flux, which if large can imply a biogenic source. In oxic atmospheres, methane abundances are controlled by the balance between surface sources and destruction via oxidation reactions with OH radicals, which in turn depend on the UV stellar spectrum (Rugheimer et al. 2015; Segura et al. 2005). However, in anoxic atmospheres such as the Archean Earth, before the advent of oxygenic photosynthesis, the $CH_4$ surface flux is approximately balanced by photolysis in the upper atmosphere by Lyman-alpha (121 nm) photons, and the rate at which $CH_4$ molecules are delivered to the upper atmosphere is, in turn, controlled by diffusion-limited hydrogen escape (Krissansen-Totton et al. 2018; Pavlov et al. 2001; Zahnle 1986). Therefore by assuming diffusion-limited escape, an inferred distribution for the minimum methane surface flux can be derived and compared to theoretical probability distributions for the maximum abiotic methane flux (e.g. Krissansen-Totton et al. 2018). Specifically, the inferred flux distribution can be repeatedly randomly sampled, and for each sampled flux value, the probability of this flux being non-biological is obtained. This probability is found by integrating the theoretical non-biological production distribution from the sampled flux value to infinity (Fig. 6). By repeating this procedure thousands of times, an average probability for the observed atmosphere being attributable by non-biological mechanisms is obtained (Fig. 6).

For our nominal Archean Earth-like 10 transit case, the probability of abiotic processes being able to explain the observed methane abundance is only 9%, although this varies considerably with different spectral noise realizations (ranging from 4% to 39% for different realizations in Fig. 3a). For 50 transits, the probability of attributing the observed methane to nonbiological processes drops to just 2%. The non-biological methane production distribution adopted from Krissansen-Totton et al. (2018) is a first attempt, and more work is needed on the geochemistry of non-biological methane production and its possible contextual clues. For example, the framework described in Fig. 6 implicitly assumed that CO and $CH_4$ outgassing scenarios have been ruled out by the non-detection of atmospheric CO. However, these calculations demonstrate that searching for biosignatures in anoxic atmospheres is feasible with JWST for TRAPPIST-1e. Furthermore, biosignature detection for TRAPPIST-1f and 1g is even more favorable than for 1e due to the lower bulk density of these planets (not shown). If TESS discovers nearby transiting, habitable planets upon which anoxic biosignatures are later detected with JWST, then an even stronger case for biology might be made when placed in a Bayesian framework for calculating the probability of life's presence (Catling et al. 2018).

Note that in our assumed distribution, the maximum non-biological methane flux (Fig. 6c) is very conservative, implying that our abiotic production probabilities may be too high. Guzmán-Marmolejo et al. (2013) argued that fluxes greater than ~1 Tmol $CH_4$/yr are difficult to explain without life, and they used a photochemical model to show non-biological methane abundances should therefore not exceed ~10 ppm. Our assumed non-biological methane flux distribution allows for higher fluxes because we allow for a broader range of crustal production rates and don't assume that $CO_2$ availability limits $CH_4$ production (Krissansen-Totton et al. 2018).

One caveat on the results described above is that heterogeneity and variability in the transit light source were not considered. It is debated whether these effects could cause large uncertainties for transit observations of late M-dwarfs. Rackham et al. (2018) argued that unocculted star spots could result in stellar contamination features in the TRAPPIST-1 transit transmission spectra that are comparable or larger in magnitude than the expected atmospheric features. Subsequently, Zhang et al. (2018) showed that this stellar model could explain recent Hubble Space Telescope (HST) and Spitzer observations of the TRAPPIST-1 planets by invoking ~30% spot coverage on the star's surface. However, a recent analysis of TRAPPIST-1 transit data from K2, SPECULOOS, Liverpool, and Spitzer telescopes ruled out the high spot coverage models of Rackham et al. (2018) and Zhang et al. (2018). Stellar models with a small coverage fraction of bright faculae are a better fit to transit data, and would not have as large an impact on planetary NIR transmission spectra (Ducrot et al. 2018; Morris et al. 2018).

Another caveat is that we optimistically assumed no noise floor and no instrumental noise other than that already prescribed in PandExo. However, we performed sensitivity tests where we repeated our Archean Earth-like retrievals with a 40 ppm noise floor and found $\log(CH_4) = -2.21^{+0.76}_{-0.93}$. This is similar to the nominal case because the largest $CH_4$ absorption features lie in regions of the spectrum where the noise level is above the noise floor. Changes in uncertainties in other atmospheric parameters are modest.

Finally, several simplifications were made in forward modeling that could affect the retrieval. For example, we assumed constant mixing ratios with altitude, which is unrealistic for some species such as $CH_4$ due to photochemical destruction at high altitude. However, self-consistent photochemical models of Archean-like atmospheres show that $CH_4$ only declines in abundance above 50-60 km (Kharecha et al. 2005; Zahnle et al. 2006), whereas the transit spectrum is mostly sensitive to abundances in the 10-60 km range (for Earth-like atmospheric structure). Nonetheless, future retrievals should be performed with self-consistent climate and photochemistry to more accurately constrain mixing ratios. Retrieved $CH_4$ abundances from models that assume constant mixing ratios with altitude should be seen as lower limits on tropospheric $CH_4$ abundances. Additionally, a photochemical model should be used to more accurately relate $CH_4$ surface fluxes to $CH_4$ mixing ratios rather than the diffusion-limited calculations adopted in Fig. 6b. This is because this relationship is $CO_2$-dependent as higher $pCO_2$ shields $CH_4$ from Lyman-alpha photons (Pavlov et al. 2001).

Although this study is focused on JWST retrievals, it is worth noting that future telescopes such as the Large-Aperture UV-Optical-Infrared (LUVOIR) mission could perform transit spectroscopy out to ~5 μm (Bolcar et al. 2016) and could therefore constrain $CO_2$, $CH_4$, and CO gas abundances much more precisely than JWST.

**4.1) Photochemical production of CO and CO anti-biosignatures**

Krissansen-Totton et al. (2018) argued that absence of CO would strengthen the $CH_4+CO_2$ disequilibrium biosignature because (i) scenarios that generate $CH_4$ abiotically such as impacts or outgassing from a strongly reduced mantle would also produce CO, and so the absence of CO would rule out these scenarios and (ii) CO is a free lunch that ought to be readily consumed by microbes, and so its persistence suggests the absence of biology. CO may also be generated photochemically from $CO_2$- dissociation and could potentially accumulate to high abundances around M-dwarfs because of the shape of the UV spectrum (Harman et al. 2015; Nava-Sedeno et al. 2016). Consequently, the absence of CO (with abundant $CH_4$ and $CO_2$) in the atmospheres of habitable planets around M-dwarfs is arguably a more compelling biosignature than around G-stars because it implies a large, presumably biological, sink for CO to balance photochemical production or other forms of abiotic production.

It could be argued that the $CH_4$ and $CO_2$ in the atmosphere of an M-dwarf planet will often be accompanied by photochemically-produced CO, and so the absence of CO is unlikely. However, on inhabited planets biological CO consumption would likely increase to draw down photochemically-produced CO to below detectable thresholds. In Appendix A, we present thermodynamic calculations demonstrating that if CO-consumers exploit the available free energy, then for the Archean Earth-like atmospheres considered in this study, the steady state CO abundance would not exceed a few ppmv, and would more likely be a few ppbv. We assume free energy-limited CO-consumption is appropriate because the only other substrate required for CO oxidation is water, which would not be limiting on habitable zone planets. Furthermore, the fluxes required to draw down worst-case scenario photochemical CO-production are only a few percent of the modern Earth's gross productivity, which suggests that nutrient availability is unlikely to limit CO consumption in most cases (see Appendix A).

In practice, CO may accumulate to somewhat higher values because biological consumption of CO in the ocean is limited by the transfer of gas across the atmosphere-ocean interface. For example, Kharecha et al. (2005) modeled the biogeochemical cycles of the Archean atmosphere and biosphere and found that CO mixing ratios could be $10^{-6}$ to $10^{-4}$ in the presence of acetogens due the limited transfer of CO across the atmosphere-ocean interface. Future work ought to incorporate CO-consumption into biogeochemical models of inhabited planets around M-dwarfs to better quantify likely CO abundances under different stellar spectra and nutrient-limitation scenarios.

**5) Conclusions**

From simulated spectra and subsequent retrievals, we conclude that the $CH_4+CO_2$ minus CO biosignature combination in anoxic atmospheres (proposed by Krissansen-Totton et al. (2018)) is potentially detectable with JWST for nearby transiting planets such as TRAPPIST-1e. For cloud-free conditions, 10 transits may be enough to constrain abundances of all three gases. The potential significance of discovering extraterrestrial biospheres means a strong case exists for searching for this disequilibrium biosignature combination on habitable exoplanets with JWST.

Oxygen-rich planetary atmospheres would take time to evolve due to the reactivity of $O_2$ and the required prior biological evolution of water-splitting photosynthesis. We have thus argued that if life exists elsewhere then the $CH_4+CO_2$ minus CO biosignature is probably much more common than the oxygen-ozone biosignature. In addition, here we have demonstrated that the biosignature combination of $CH_4+CO_2$ minus CO in anoxic, Archean-like atmospheres will be easier to detect than ozone with JWST.

Retrieved posterior probability distributions of $CH_4$ abundances can be combined with theoretical calculations of maximum non-biological methane production to calculate the probability that the observed $CH_4$ can be explained by non-biological processes. We find that JWST observations could be used to make quantitative inferences about the chance of the data being attributable to extraterrestrial biology. For Archean Earth-like methane levels on TRAPPIST-1e, we make a preliminary estimate that for 10 transits the probability of abiotic processes being responsible for the methane would be ~9%. For 50 transits, the probability drops to 2%. Moreover, we note that there is debate about whether the efficacy of abiotic methane production has been overestimated (see Krissansen-Totton et al. (2018)), and so these probabilities for abiotic explanations likely err on the conservative side.

**Acknowledgements**

We thank Jean-Loup Baudino, Sophie Bauduin, Andrew Lincowski, Jacob Lustig-Yaeger, Ryan MacDonald, Daniel Toledo and the anonymous reviewer for helpful discussions and suggestions that greatly improved the manuscript. J.K.-T. is supported by NASA Headquarters under the NASA Earth and Space Science Fellowship program (grant NNX15AR63H). This work was also supported by the NASA Astrobiology Institute's Virtual Planetary Laboratory (grant NNA13AA93A) and the cross-campus Astrobiology Program at the Univ. of Washington.

**Appendix A: Energy-limited carbon monoxide consumption**

This section calculates likely steady-state CO abundances in Archean Earth-like atmospheres assuming biological CO consumption uses the available free energy. The two relevant metabolisms are specified by the following equations:

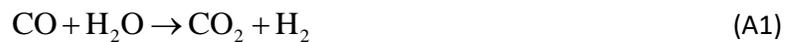

$$\mathrm{CO} + \mathrm{H_2O} \rightarrow \mathrm{CO_2} + \mathrm{H_2} \quad \text{(A1)}$$

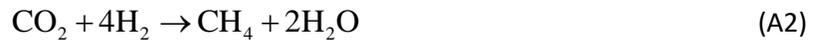

$$\mathrm{CO_2} + 4\mathrm{H_2} \rightarrow \mathrm{CH_4} + 2\mathrm{H_2O} \quad \text{(A2)}$$

Since we are considering planets with $CH_4$+$CO_2$ biosignatures, methanogenesis (equation (A2)) is present by assumption. The net result of these two metabolisms is:

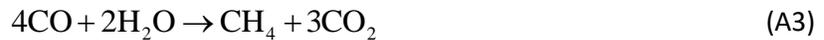

$$4\mathrm{CO} + 2\mathrm{H_2O} \rightarrow \mathrm{CH_4} + 3\mathrm{CO_2} \quad \text{(A3)}$$

Given this net reaction and assumed $CH_4$ and $CO_2$ abundances, the CO abundance at which the net biological reaction proceeds at the limit of thermodynamic viability can be calculated. The biosphere will draw down atmospheric CO to this level assuming metabolic activity is not limited by the availability of other substrates or nutrients (see below).

Following Kral et al. (1998) and Kasting et al. (2001), we conservatively assume the reaction (A3) is no longer thermodynamically viable when the Gibbs energy yield equals the Gibbs energy required to synthesize four moles of ATP (4 moles of CO are oxidized in reaction (A1)). In other words, we are solving the following equation:

$$\Delta G_{4\,ATP} = \Delta G_0 + RT \ln\left( \frac{p\mathrm{CH_4} \times (p\mathrm{CO_2})^3}{(p\mathrm{CO})^4} \right) \quad \text{(A4)}$$

Here, $\Delta G_{4\,ATP}$ =-142 kJ/mol is the change in Gibbs energy required to synthesize 4 mol of ATP, $\Delta G_0$ = -227.2 kJ/mol is the Gibbs energy of the reaction at standard conditions (calculated using code and databases described in Krissansen-Totton et al. (2016)), R=8.314 J/mol/K is the gas constant, T=300 K is surface temperature, and $p\mathrm{CH_4}$, $p\mathrm{CO_2}$, and $p\mathrm{CO}$ are gas partial pressures (in bar). Note that the activity of water is unity since water is not supply limited at the surface of habitable planets. For the Archean Earth-like planets in our paper $p\mathrm{CH_4}$=0.005 bar and $p\mathrm{CO_2}$=0.05 bar. Solving this equation yields pCO =

6×10$^{-6}$ bar. This is almost certainly an overestimate of the energy-limited CO abundance because chemoautotrophic metabolisms are known to metabolize at free energies less than the amount required for ATP synthesis (Conrad 1996). For example, methanogenesis can be supported by less than half the Gibbs energy yield from ATP synthesis (Conrad 1996), and so adopting a more realistic value of $\Delta G_{2ATP}$ on the left hand side of equation (A4) yields pCO = 5×10$^{-9}$ bar.

Is it reasonable to assume biological CO-consumption would be energy limited? Harman et al. (2015) modeled the photochemical production of CO in Earth-like atmospheres and found steady-state CO surface deposition fluxes of up to 5×10$^{11}$ molecules/cm$^2$/s (1.3×10$^{14}$ mol C/yr) were required to balance photochemical production around M-dwarfs. The carbon throughput on Earth's terrestrial biosphere is around 10$^{16}$ mol/yr (Beer et al. 2010), whereas the primary ocean productivity is around 4×10$^{15}$ mol C/yr (del Giorgio & Duarte 2002). Consequently, any CO-consuming biosphere need only be a few percent as productive as Earth's biosphere to draw down atmospheric CO. The only other substrate in the CO-consuming reaction, water, would not be limiting on the surfaces of habitable planets by assumption. Biospheres may exist that are so severely nutrient-limited that photochemically-produced CO accumulates despite CO consumption. However, based on the above considerations we expect biological CO drawdown to be the norm rather than the exception on inhabited worlds.

**Appendix B: Alternative radius formulation and radius-abundance degeneracy**

In the main text the planet radius parameter used in our retrievals was defined as the solid-body radius. Here, we repeat our nominal Archean Earth-like NIRSpec prism retrieval (Fig. 2) where the radius parameter now represents the 1 mbar radius. This approach is more typical of retrievals of giant planet atmospheres, and it ensures radius and surface pressure are independent. By defining the radius as the 1 mbar radius the degeneracy between gas abundances and radius are more clearly revealed than when using surface radius, which is anticorrelated with surface pressure (Fig. 2).

The results are shown in Fig. A1. The joint distributions for radius and gas abundances now show a tight anticorrelation because if the the 1 mbar radius is increased then the abundances of absorbing gases must be decreased to produce the same transit heights. However, the marginal distributions for gas abundances are virtually identical to Fig. 2 because we are merely changing the basis vector that describes our atmosphere.

**Figures:**

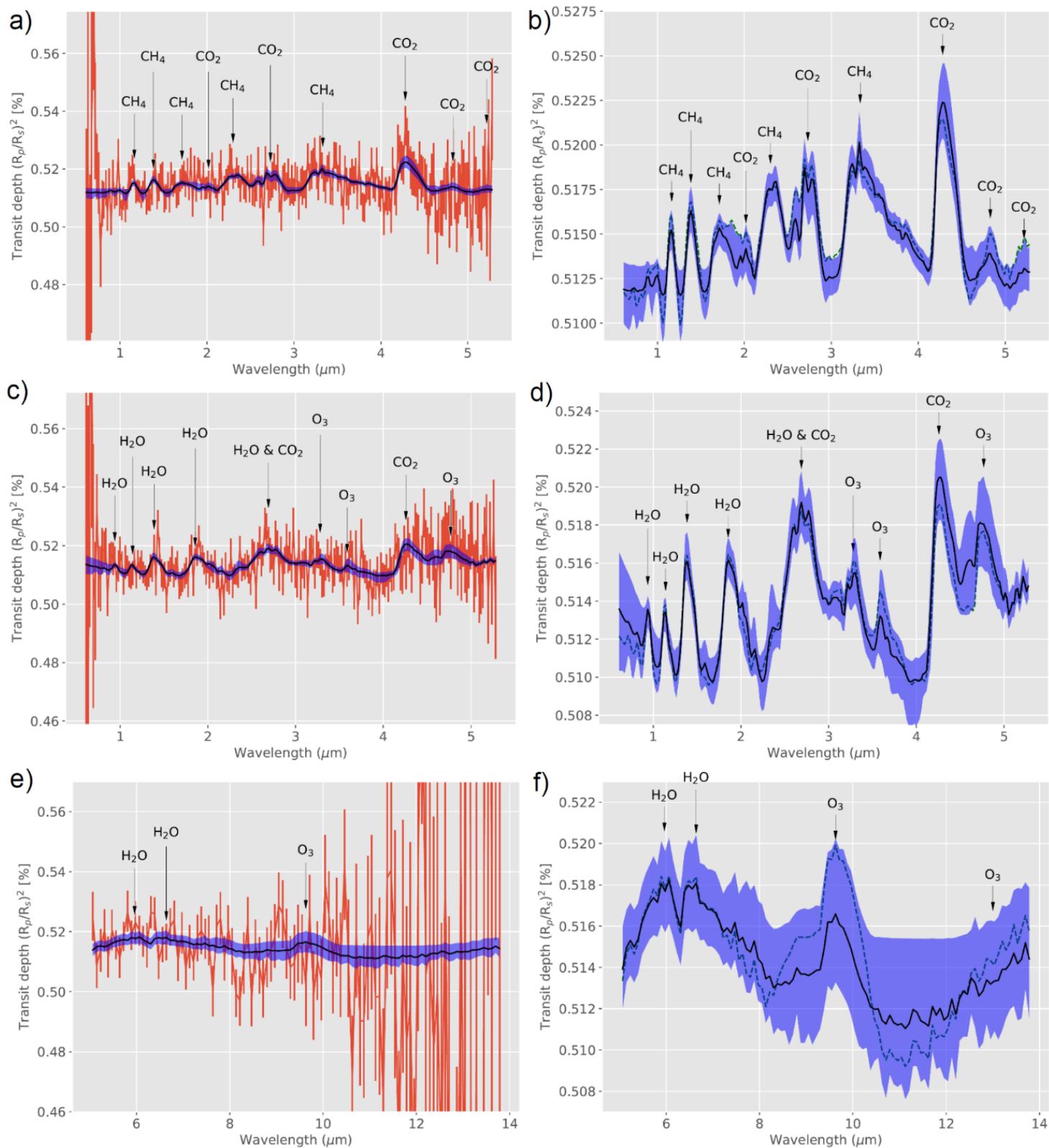

**Fig. 1**: Synthetic and fitted spectra for 10 transits of TRAPPIST-1e with no clouds where $R_P$ and $R_S$ are the radii of the planet and star, respectively. Left-side panels show typical noise realizations using PandExo (red lines), the median fitted spectrum calculated using the Nested Sampling retrieval algorithm (black lines) with 95% credible intervals from the retrieval (blue shaded regions). The right-side panels show the same median fit and 95% credible intervals, in addition to the true synthetic spectrum (green-dashed line). The right-side panels have a smaller y-axis range such that individual spectral features can be more easily seen. Top row shows the Archean Earth-like case using NIRSpec prism, the middle row shows the modern Earth-like case using NIRSpec prism, and bottom row shows the modern Earth-like case using MIRI LRS. Key molecular absorption features are labeled. Note that stratospheric water vapor abundances are assumed to be unrealistically high to maximize the possible obscuration of $CH_4$ and $CO_2$ features (see main text).

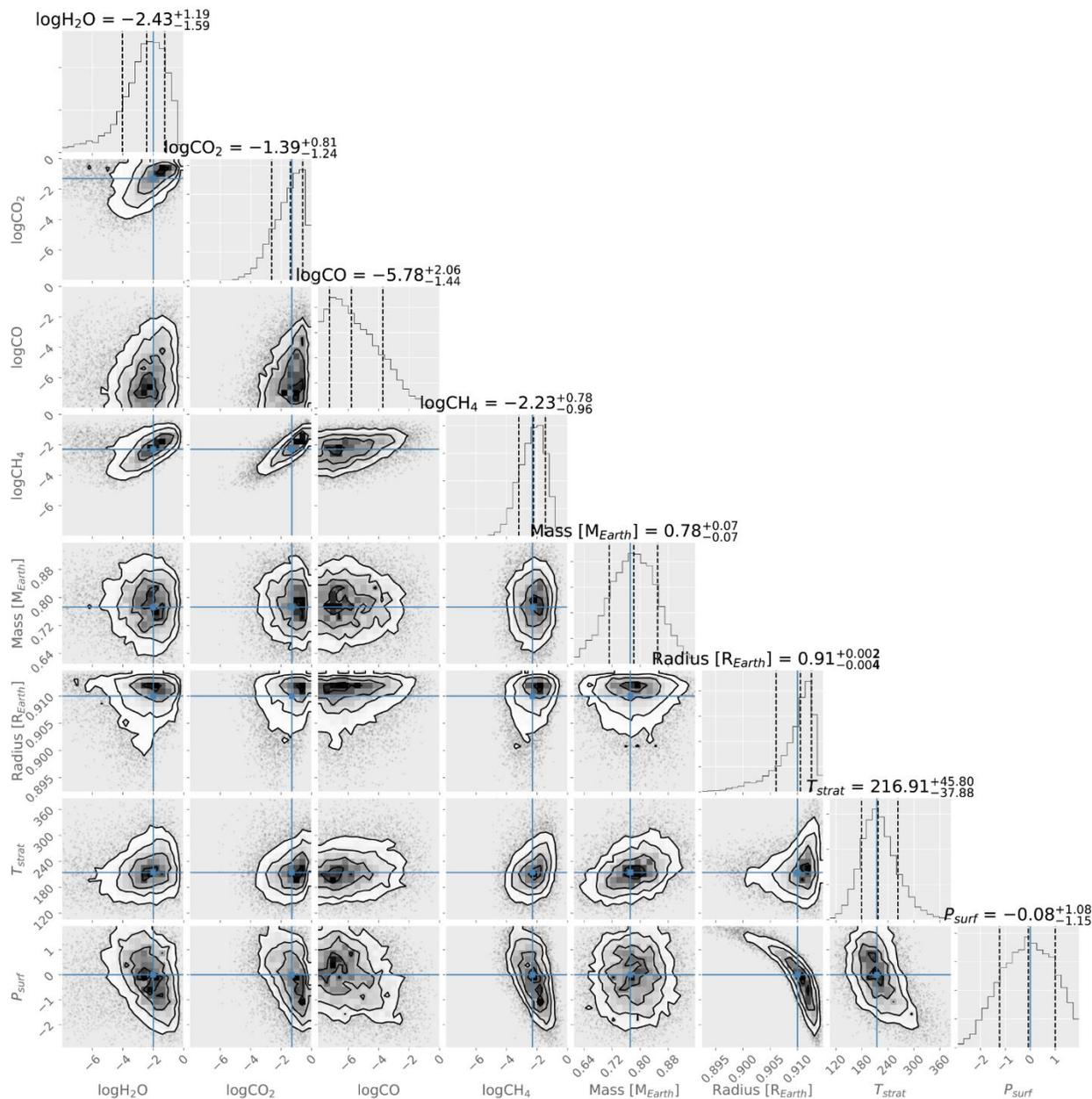

**Fig. 2**: Posterior probability distributions for retrieved parameters for 10 transits of an Archean Earth-like TRAPPIST-1e with no clouds using NIRSpec prism produced using the plotting script corner.py (Foreman-Mackey 2016). Diagonal elements are marginal distributions, off-diagonal elements are joint distributions, and vertical and horizontal blue lines are "true" values. The marginal distributions show that $CH_4$ and $CO_2$ are detectable, and that $CH_4$ abundances can be constrained to within a few orders of magnitude. It is also possible to put tentative upper bound on CO abundance. For this retrieval, the midpoint of all spectral data points was the true spectrum to ensure posteriors were centered on true values, but note that individual noise realizations may vary (Feng et al. 2018) (Fig. 3a).

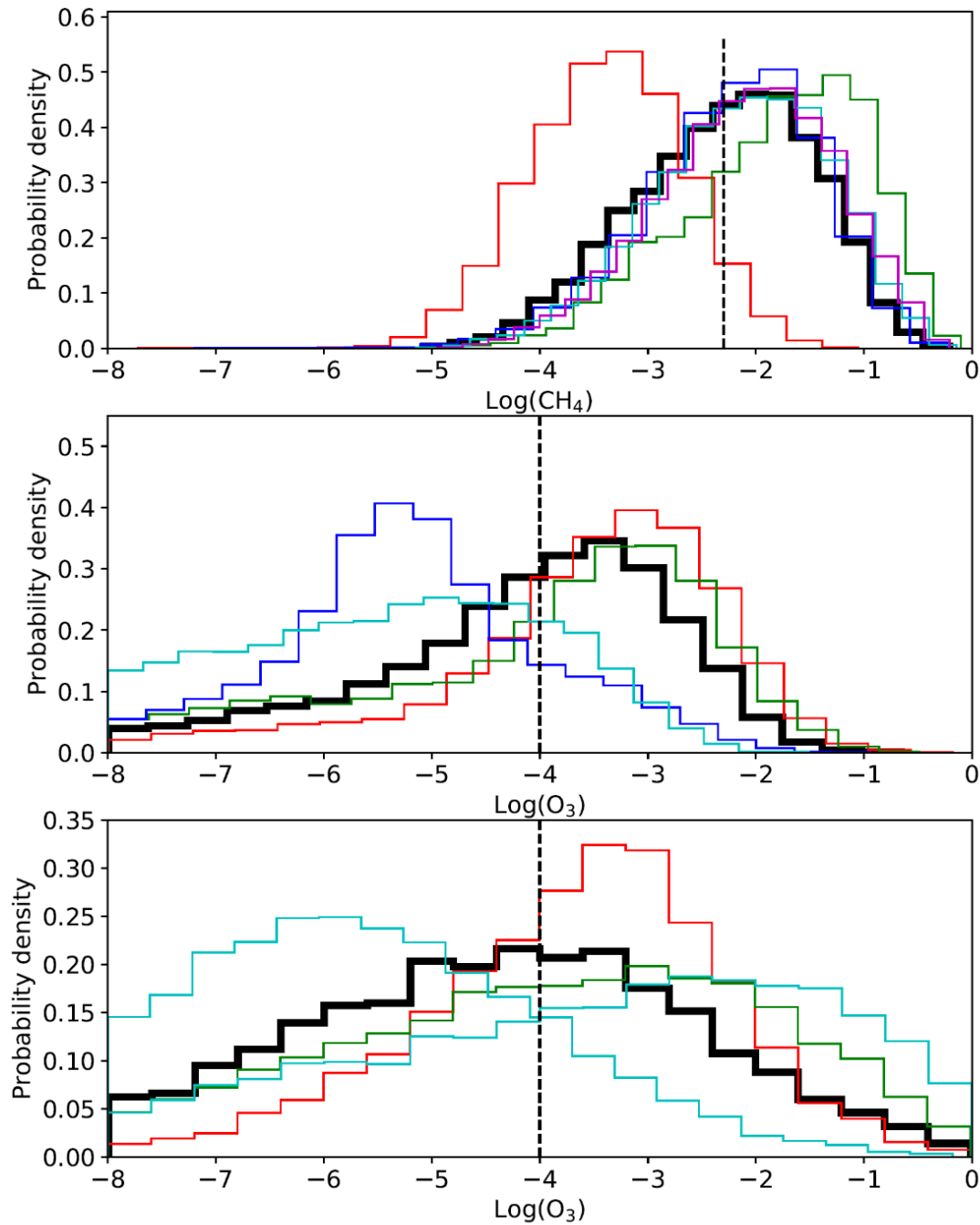

**Fig. 3**: Comparison of (a) $CH_4$ posteriors from Archean Earth-like NIRSpec prism case, (b) $O_3$ posteriors from modern Earth-like NIRSpec prism case, and (c) $O_3$ posteriors from modern Earth-like MIRI LRS case (all three are for 10 coadded transits of TRAPPIST-1e with no clouds). Bold black lines show posteriors for noise realizations centered on true values (Feng et al. 2018), whereas thin colored lines show randomized noise realizations (see main text). Vertical black dashed lines denote the "true" parameter values. It is possible to detect and constrain Archean-like biogenic $CH_4$ abundances with NIRSpec prism, whereas for the same number of transits, $O_3$ detection is not possible with either NIRSpec or MIRI. Note that $O_3$ mixing ratios of $10^{-4}$ are far larger than what would realistically be expected on an inhabited planet, and so $O_3$ detection would likely be even more challenging than (b) and (c) imply.

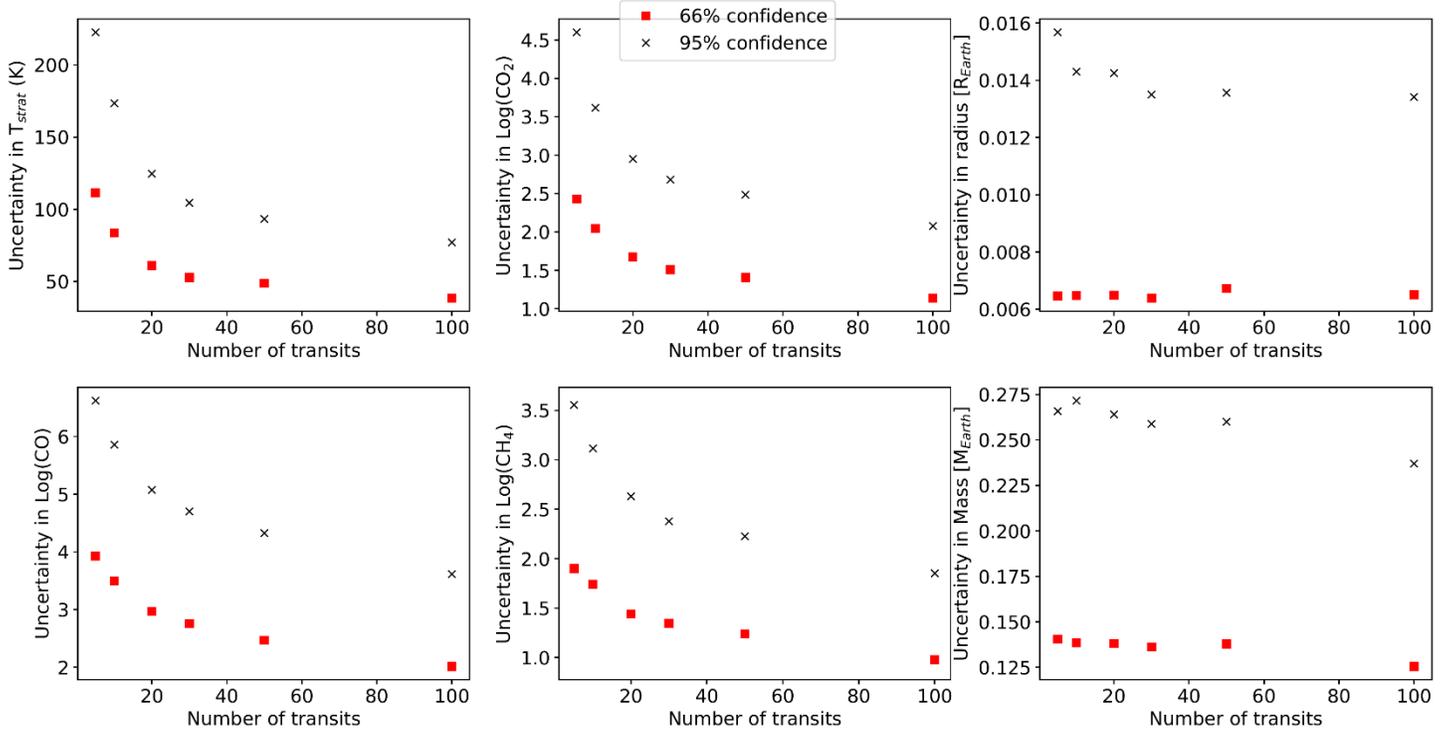

**Fig. 4**: Uncertainty in parameter posteriors as a function of number of transits. Red squares and black crosses show 66% and 95% credible interval uncertainties, respectively. These results are in general agreement with those of Batalha et al. (2018) which show diminishing returns with more transits. However, 30-50 transits would reduce uncertainty in $CH_4$ compared to the 10 transit case, and therefore allow a stronger inference to biology (see discussion). For all the retrievals plotted above, the midpoint of all spectral data points was the true spectrum to reduce stochastic variation in posteriors (Feng et al. 2018).

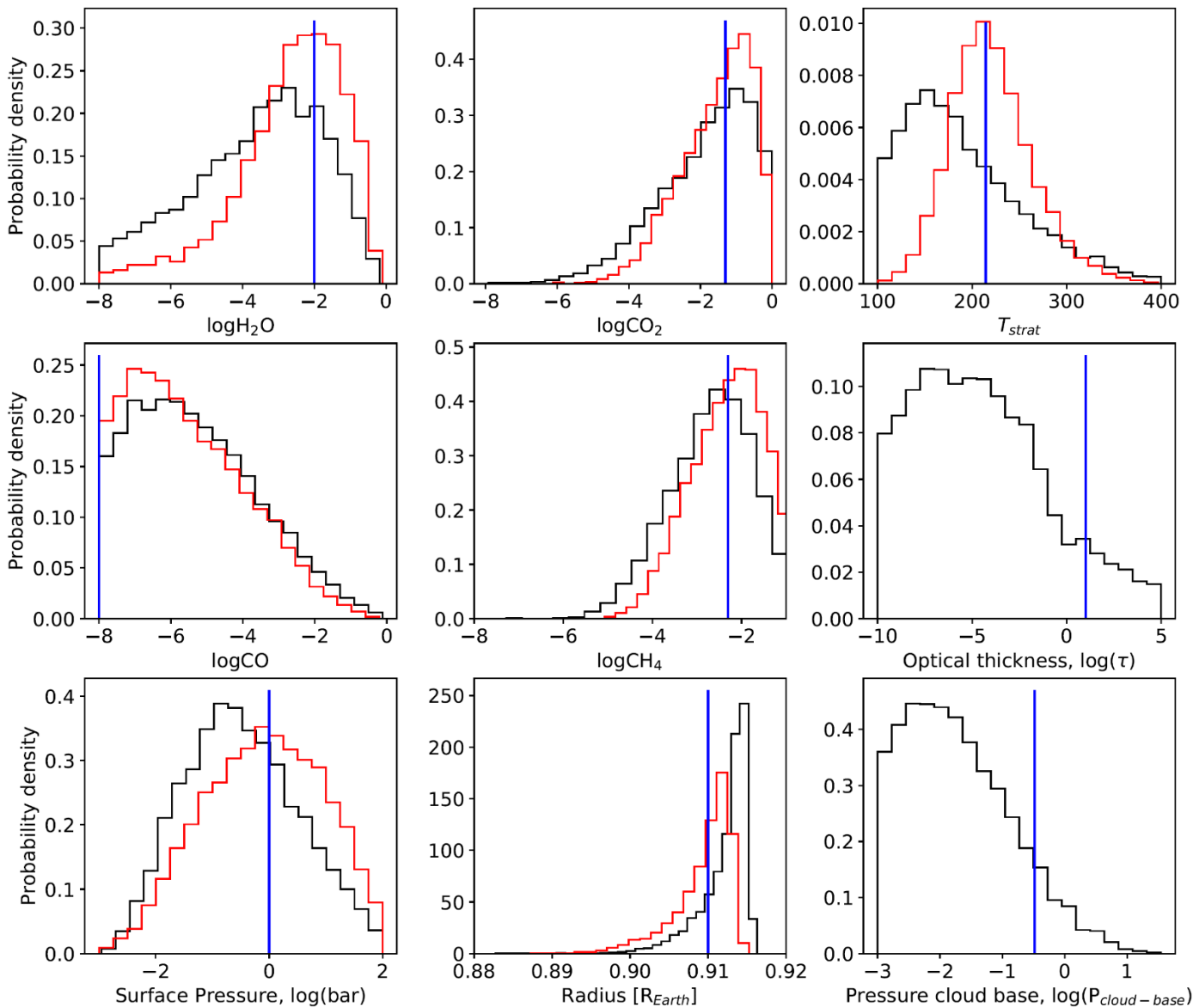

**Fig. 5**: Posterior probability distributions for selected parameters for 10 transits of an Archean Earth-like TRAPPIST-1e with high clouds using NIRSpec prism. Black lines denote posterior probability distributions for the cloudy case, red lines denote posterior distributions for the no-cloud Archean case (Fig. 2) for comparison, and blue vertical lines denote "true" values. Clouds widen the posterior probability distributions for gas abundances somewhat, but even for this high cloud case where the transmission spectrum is truncated at ~30 km (0.01 bar), $CO_2$ and $CH_4$ detection is possible. Posteriors for fractional scale height and planet mass are not plotted.

18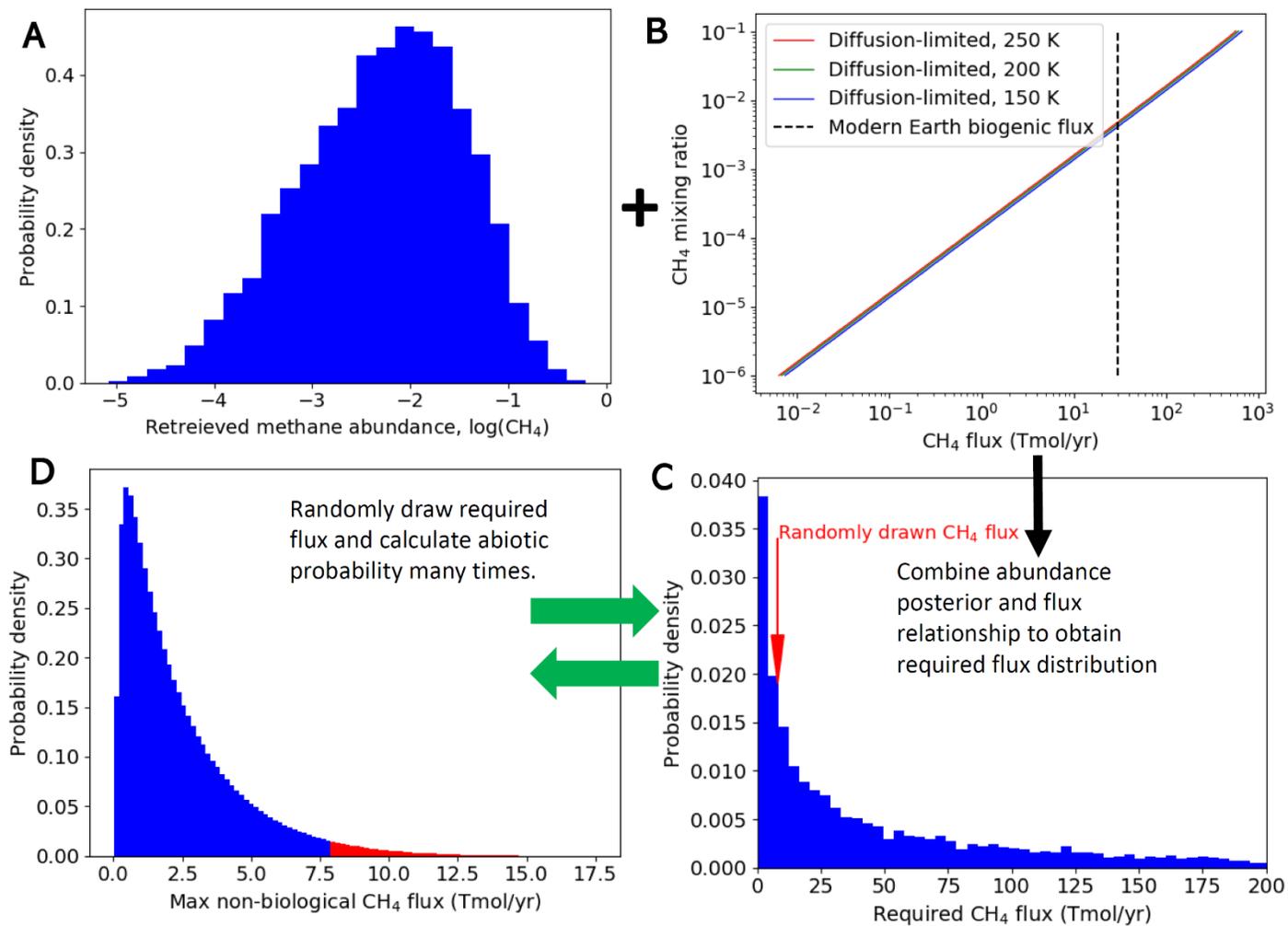

**Fig. 6:** One approach to making probabilistic inferences to biology from atmospheric abundance constraints. The posterior distribution for CH$_4$ (A) is converted to a necessary replenishing surface flux distribution (C) by assuming diffusion-limited escape (B, adapted from Krissansen-Totton et al. (2018)). This distribution for the required flux (C) is then repeatedly sampled, and each sampled value is compared to a theoretical distribution for the maximum non-biological methane flux (D, adapted from Krissansen-Totton et al. (2018)). Specifically, the probability of the sampled flux occurring via non-biological processes is found by integrating the (D) from the sampled flux value to infinity. For example, if 8 Tmol/yr is drawn from (C, red arrow), then the probability of this being nonbiological is obtained by integrating (D) from 8 Tmol/yr to infinity (red color region). This is repeated 10,000 times, and the average of these values is the probability that the observed methane abundance can be explained by non-biological processes.

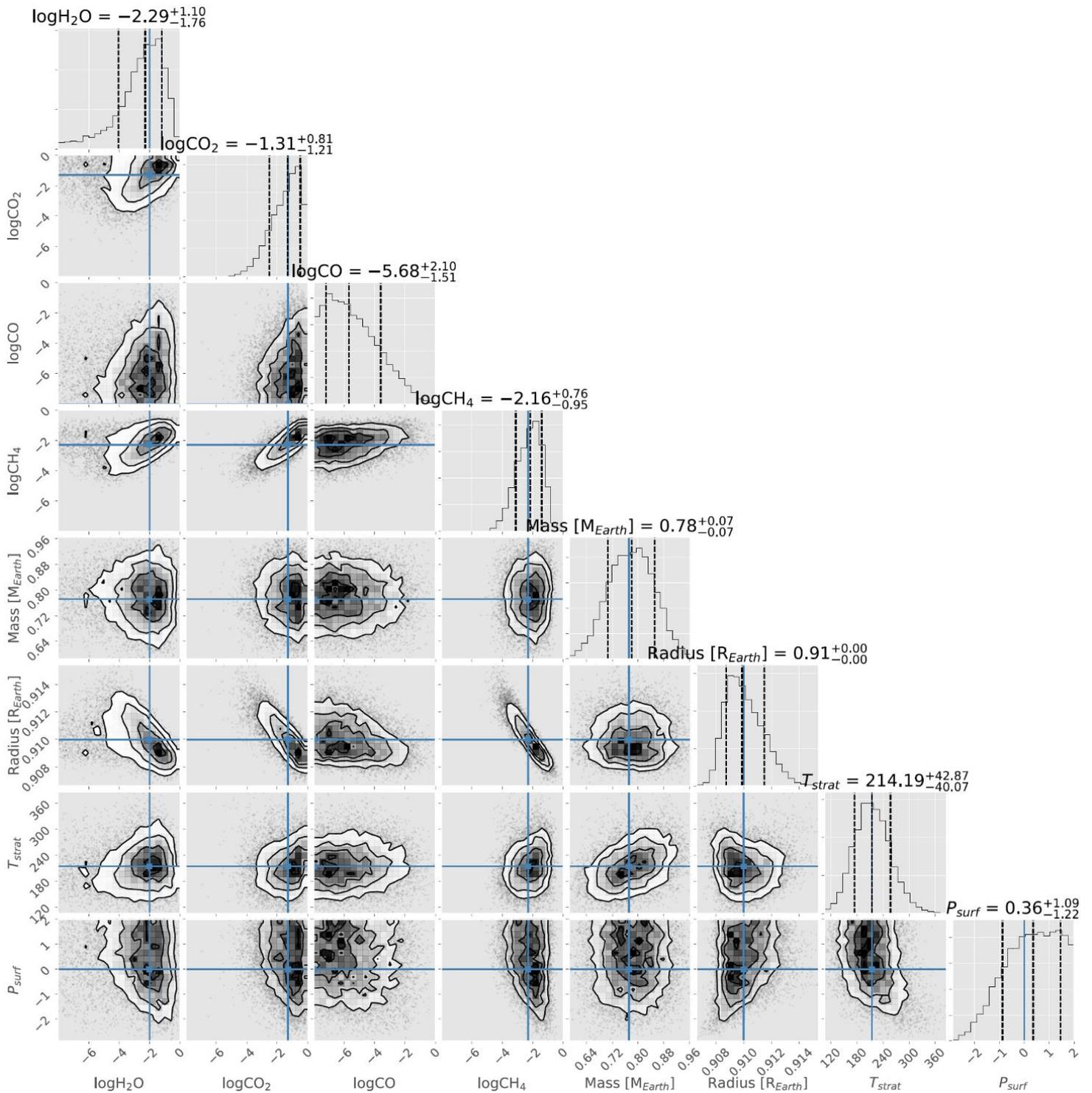

**Fig. A1**: Same as Fig. 2 except in this retrieval the planet radius parameter represents the 1 mbar radius rather than the solid-body (surface) radius. The joint distributions between radius and gas abundances now show a clear anti-correlation, but the marginal distributions for gas abundances are unchanged.